\documentclass[reprint,
superscriptaddress,
amsmath,amssymb,
aps,
pra,
balancelastpage]{revtex4-1}

\usepackage{graphicx}
\usepackage{dcolumn}
\usepackage{bm}
\usepackage{hyperref}
\usepackage[dvipsnames]{xcolor}

\begin{document}


\title{Unveiling the electron-nuclear spin dynamics in an \emph{n}-doped InGaAs epilayer\\
by spin noise spectroscopy}

\author{C.~Rittmann}
\affiliation{Experimentelle Physik 2, Technische Universit\"at Dortmund, 44221 Dortmund, Germany}

\author{M.~Yu.~Petrov}
\affiliation{Spin Optics Laboratory, Saint Petersburg State University, 198504 St.\,Petersburg, Russia}

\author{A.~N.~Kamenskii}
\affiliation{Experimentelle Physik 2, Technische Universit\"at Dortmund, 44221 Dortmund, Germany}

\author{K.~V.~Kavokin}
\affiliation{Spin Optics Laboratory, Saint Petersburg State University, 198504 St.\,Petersburg, Russia}

\author{A.~Yu.~Kuntsevich}
\affiliation{P.~N. Lebedev Physical Institute, Russian Academy of Science, 119991 Moscow, Russia}

\author{Yu.~P.~Efimov}
\author{S.~A.\ Eliseev}
\affiliation{Resource Center Nanophotonics, Saint Petersburg State University, 199034 St.\,Petersburg, Russia}

\author{M.~Bayer}
\affiliation{Experimentelle Physik 2, Technische Universit\"at Dortmund, 44221 Dortmund, Germany}

\author{A.~Greilich}
\affiliation{Experimentelle Physik 2, Technische Universit\"at Dortmund, 44221 Dortmund, Germany}

\date{\today}

\begin{abstract}
We discuss the implications of a small indium content (3\%) in a GaAs epilayer on the electron- and nuclear-spin relaxation due to enhanced quadrupolar effects induced by the strain. Using the weakly perturbative spin-noise spectroscopy, we study the electron-spin relaxation dynamics without explicit excitation. The observed temperature dependence indicates the presence of localized states, which have an increased interaction with the surrounding nuclear spins. Time-resolved spin-noise spectroscopy is then applied to study the relaxation dynamics of the optically pumped nuclear-spin system. It shows a multi-exponential decay with time components, ranging from several seconds to hundreds of seconds. Further, we provide a measurement of the local magnetic field acting between the nuclear spins and discover a strong contribution of quadrupole effects. Finally, we apply the nuclear spin diffusion model, that allows us to estimate the concentration of the localized carrier states and to determine the nuclear spin diffusion constant characteristic for this system.
\end{abstract}

\maketitle


\section{\label{sec:introduction}Introduction}

Silicon dopants (Si) are known to provide shallow donors in gallium arsenide (GaAs)~\cite{Brozel}. Depending on the crystal quality and the concentration of such impurities in $n$-doped GaAs (GaAs:Si), the electrons can show extremely long times of spin relaxation~\cite{KikkawaPRL98, DzhioevPSS97, DzhioevPRB02, ColtonPRB04, FuPRB06, KavokinSST08, BelykhPRB17, LonnemannPRB17} limited by the Dyakonov-Perel mechanism~\cite{DP_PSS71, DP_JETP71} or spin-spin exchange interactions~\cite{KavokinPRB01}, and even experience weak localization~\cite{BelykhPRX18} at high and moderate concentrations of dopants.
At relatively small doping density, $n_D \lesssim 10^{15}$\,cm$^{-3}$, well below the metal-insulator transition (MIT)~\cite{mott1968metal}, at which the motion-assisted and spin-exchange relaxation mechanisms are suppressed, the hyperfine interaction of a localized-electron spin with the surrounding spins of lattice nuclei provides the dominant mechanism of spin relaxation~\cite{DzhioevPRL02}.
In addition, since each electron is localized on a shallow impurity, the regime of short correlation time is realized even at liquid-helium temperature~\cite{SokolovPRB17}.
On the contrary, in singly-charged quantum dots (QDs), where each electron is strongly localized due to the spatial confinement and the reduced bandgap of Indium-Gallium Arsenide (InGaAs), the correlation time is long~\cite{MerkulovPRB10}.
In this case, the precession of an electron spin around the randomly oriented field of nuclear spin fluctuations results in a fast dephasing of the electron spin orientation, during a nanosecond time scale, followed by a total loss of the spin coherence during a microsecond time scale~\cite{MerkulovPRB02, KhaetskiiPRL02}.
\looseness=+1

Research on spin systems in the context of quantum information processing and data storage is guided by the need of a  technical implementation of a largely isolated nuclear-spin system with well controlled nuclear state and reduced fluctuations.
Such a control can be achieved when the phase transition to a spin-ordered state is realized~\cite{MerkulovJETP82, MerkulovPSS98, KorenevPRL07, ScalbertPRB17}. However, nuclear spin ordering has not yet been observed since it requires extremely small nuclear-spin temperatures $\sim$\,0.1\,$\mu$K~\cite{VladimirovaPRB21}. Recent achievements in nuclear-spin cooling in GaAs:Si have shown values at least an order of magnitude larger than the theoretical estimations demand~\cite{KKF82, VladimirovaPRB18, Kotur21, Litvyak21}.
On the other hand, the nuclear spin-relaxation times observed in InGaAs/GaAs QDs were found to be longer than in lattice-matched AlGaAs/GaAs QDs due to enhanced quadrupole interactions, suppressing the nuclear spin fluctuations~\cite{DzhioevPRL07, makhoninFastControlNuclear2011, chekhovich2015suppression}.
Generally, the nuclear spin system of InGaAs is more complex than that of GaAs due to the additional indium content.
Strong quadrupole effects are observed even in highly annealed QD ensembles~\cite{FlisinskiPRB10, SokolovPRB16} for which the concentration of indium is reduced.
\looseness=+2

In this paper, we study both the electron- and nuclear-spin dynamics in an epilayer of InGaAs. The doping by Si provides centers of electron localization similar to GaAs:Si. In addition, due to the admixture of the indium, the bandgap of epilayer material is shifted to the transparency region of the GaAs substrate. For studying the electron-nuclear spin dynamics, this allows us to implement the spin-noise spectroscopy~\cite{zapasskii2013spin} which is an especially powerful method for characterizing spin systems (almost) nonperturbatively~\cite{oestreich2005spin, crooker2009spin}.
As we will show below, the electron-spin dynamics reveals the behavior of well-localized centers, similar to ensembles of singly-charged QDs, even at moderate Si doping. At the same time, the dynamics of the nuclear spin polarization is similar to GaAs:Si but characterized by complex relaxation processes assisted by the strong quadrupole interaction.
\looseness=+1

\section{\label{sec:experiment}Sample and setup}

We use a $d=10$\,$\mu$m thick InGaAs epilayer with 3\% of indium content grown by molecular beam epitaxy on a GaAs substrate. As the comparatively large indium atoms (atomic number: $Z = 49$, nuclear spin: $I = 9/2$) replace smaller gallium atoms ($Z = 31$, $I = 3/2$), the band structure is gently tuned. The sample was doped homogeneously by Si-atoms with a density of $n_D = 3.1 \times 10^{16}$\,cm$^{-3}$, measured by a 4-point method at temperature $T = 77$\,K in a magnetic field $B = 0.5\,$T. This concentration was found to maximize the electron-spin relaxation time compared to similar samples with smaller doping concentrations~\cite{Evdokimov_2019}.

\begin{figure}[t]
\includegraphics[width=.95\columnwidth]{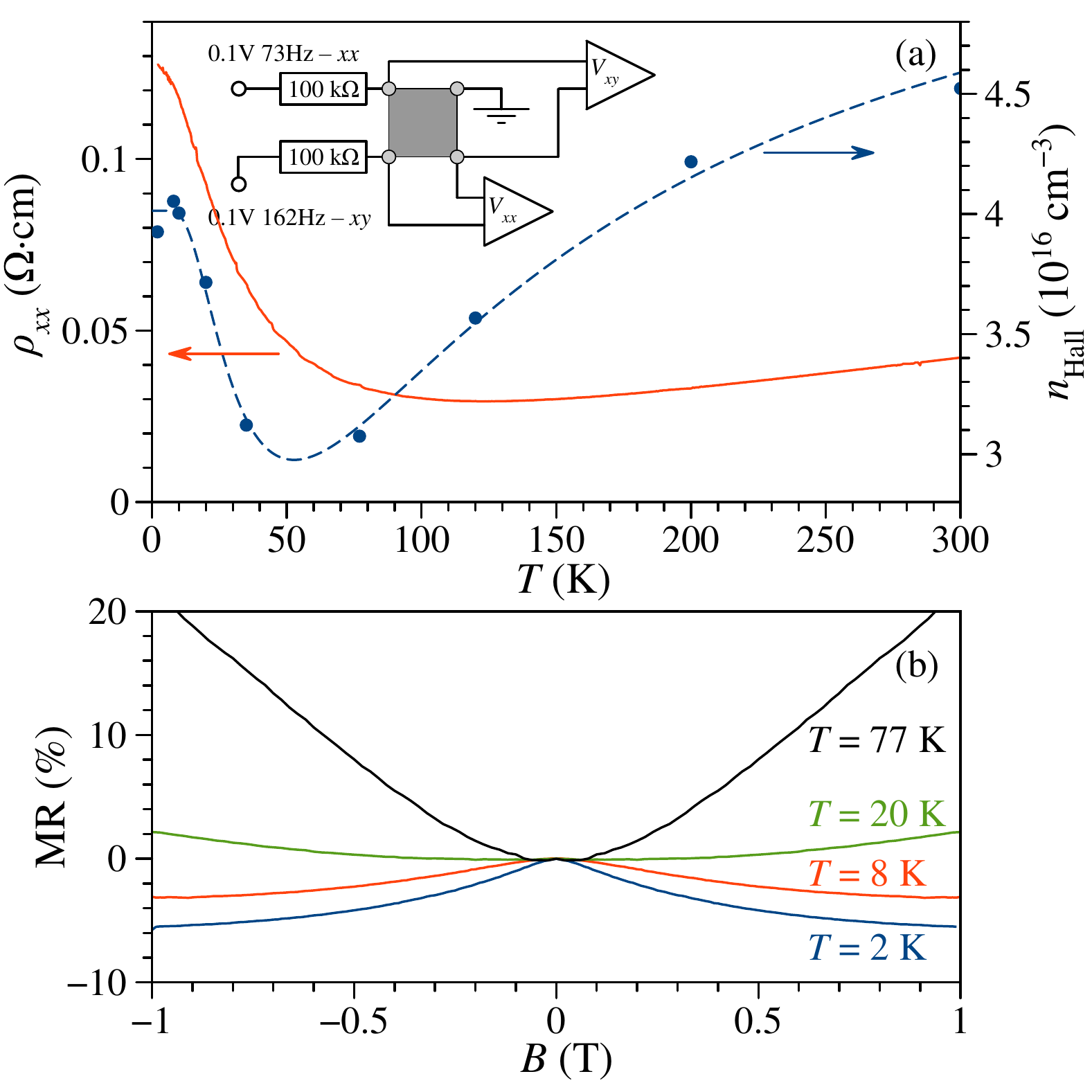}
\caption{(a) Temperature dependencies of the low-field resistivity (solid line) and the Hall carrier density (points) with the dashed line as guide to the eye. Inset shows the schematic of the transport experiment. (b) Magnetoresistance curves measured at different temperatures.}
\label{fig:Hall}
\end{figure}

As the carrier concentration plays an important role for further considerations, we performed additional transport measurements in the Van der Pauw geometry. Simultaneous measurements of the Hall effect and resistivity were performed using two lock-ins with different frequencies ($73$ and $162$\,Hz respectively) for a transport current $J = 1$\,$\mu$A, as shown in the insert of Fig.~\ref{fig:Hall}(a). The lock-in amplifiers were checked to avoid interference effects and the transport current was proven to not overheat the system at the lowest temperature. We could vary the temperature from $T = 2$ to $300$\,K and apply a magnetic field up to $8$\,T. The magnetic field perpendicular to the wafer plane was swept from positive to negative values and the longitudinal resistance $R_{xx}$ and the Hall resistance $R_{xy}$ data were symmetrized and antisymmetrized, correspondingly, to compensate for imperfect contact alignment leading to the mutual admixture of $R_{xx}$ and $R_{xy}$.

The temperature dependence of the sheet resistivity, $\rho_{xx}$, measured for $B=0$\,T and the carrier concentration extracted from the Hall effect via $V_{xy} = JB/(q_e n_\mathrm{Hall} d)$ are shown in Fig.~\ref{fig:Hall}(a). As one can see from the figure, the resistivity drops and saturates with increasing sample temperature while the concentration of conducting electrons exhibits a minimum at temperature $T\simeq 50$\,K.

To further investigate the carrier mobility, we have performed measurements of the magnetoresistance, $\mathrm{MR}$, and the Hall resistivity. The extracted $\mathrm{MR}$ in low magnetic fields was measured at several temperatures, as shown in Fig.~\ref{fig:Hall}(b). The temperature and magnetic-field dependencies of the resistivity are typical for GaAs-based semiconductors with a doping level above the MIT~\cite{Woods64, Halbo68, Capoen93}, in particular, with respect to the low-$T$ resistivity upturn, the weak localization-caused negative $\mathrm{MR}$ at low $T$ and the positive $\mathrm{MR}$ at higher temperatures. The low-$T$ value of the resistance ($\sim\,40$\,$\Omega$) and the Hall slope ($16$\,$\Omega$/T) imply that only a $d=10$\,$\mu$m thick epilayer contributes to the conductivity, at least at low temperatures. The Hall carrier density saturates at the level of $n_0 = 3.9 \times 10^{16}$\,cm$^{-3}$, slightly larger than the one measured previously for a larger $T$ ($3.1 \times 10^{16}$\,cm$^{-3}$)~\cite{Evdokimov_2019}.

\begin{figure}[t]
\includegraphics[width=.95\columnwidth]{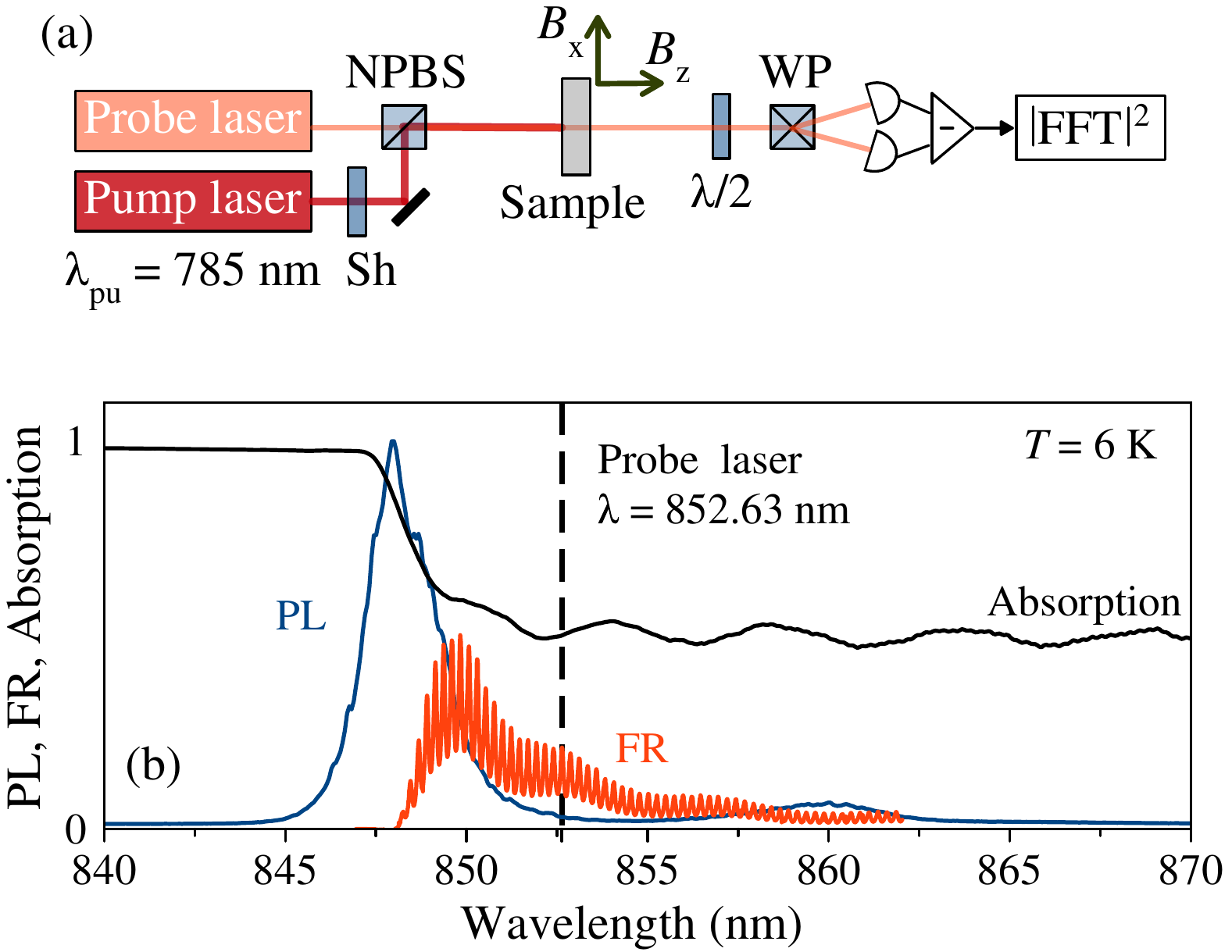}
\caption{\label{fig:setup} (a) Schematic of the setup used to detect spin noise spectra by measuring FR with the probe beam and to polarize the spin system with the pump beam. Sh is the shutter used to switch on and off the pump, and NPBS is a non-polarizing beam splitter, used to combine the linearly polarized probe and circularly polarized pump. (b) Normalized  photoluminescence (PL) (blue line), Faraday rotation (red line), and absorption (black line) spectra of the InGaAs epilayer. Vertical dashed line is the position of the probe laser.
}
\end{figure}

Figure~\ref{fig:setup}(a) schematically shows the setup for the studies of Faraday rotation (FR), spin-noise spectroscopy (SNS), and time-resolved SNS with optical pumping. The average electron-spin polarization along the $z$-direction (optical axis) is detected through the FR measured by a linearly polarized continuous wave (CW) laser, referred to as the probe laser. We fix the probe laser wavelength at $\lambda_{\text{pr}} = 852.63$\,nm, which is 4.63\,nm above the maximum of the photoluminescence signal of the InGaAs sample ($\lambda_{\text{PL,max}}$(6\,K) = 848\,nm), and hence, propagates with reduced absorption through the sample at the local maximum of the Faraday rotation (see below), as illustrated in Fig.~\ref{fig:setup}(b). Furthermore, the sample was polished from the back side to allow for transmission measurements and covered by an antireflection coating from both sides to reduce the etalon effects of the laser. Still, some residual effect was observed in the Faraday rotation measurements, see the oscillating red line in Fig.~\ref{fig:setup}(b).

Behind the sample, the induced FR of the probe beam is detected using a half-wave plate followed by a Wollaston prism and a balanced photoreceiver. The differential signal is digitized and fast Fourier transformed by a field-programmable gate array, calculating the power spectral density (PSD) of the measured signal~\cite{crooker2010QD}. The bandwidth of the diodes determines the covered spectral range and is limited to 100\,MHz in our case (Femto HCA-S).

To extract the carrier spin-noise from the background electronic and shot noise, the measured power spectrum $S_{\text{meas}}(f)$ is subtracted and divided by a reference signal $S_{\text{ref}}(f)$, which is recorded at a different external magnetic field ($B_x=14$\,mT). The resulting spin noise is expressed in shot-noise units (SNU) as $\mathrm{PSD(SNU)}=S_{\text{meas}}(f) / S_{\text{ref}}(f) - 1$.

An additional circularly polarized CW pump laser at $\lambda_{\text{pu}} = 785$\,nm is used to create a non-zero electron-spin polarization. We apply it in order to measure the luminescence and conventional Faraday rotation spectra, shown in Fig.~\ref{fig:setup}(b)~\cite{zapasskii2013optical}. Due to the high energy excitation (above the band gap), the pump-laser beam is absorbed by the sample and does not pass to the detection channel. This allows us to use a collinear scheme, in which pump and probe take the same optical path.

If not stated differently, the powers of the laser beams are $P_{\text{pr}} = 1$\,mW for the probe and $P_{\text{pu}} = 0.3$\,mW for the pump. Both beams are tightly focused onto the sample surface with spot diameters of $D_{\text{pr}} \approx 13$\,$\mu$m and $D_{\text{pu}} \approx 40$\,$\mu$m. The pump beam can be blocked by a remotely controlled shutter with a switching time resolution of $\sim$10\,ms.

The InGaAs epilayer is mounted in the center of two pairs of electromagnetic coils generating corresponding magnetic fields, $B_\text{x}$ and $B_\text{z}$, see Fig.~\ref{fig:setup}(a). The sample is placed into a helium flow cryostat at a temperature of $T=6$\,K.

\section{\label{sec:electronic_properties}Electron-spin dynamics}

\begin{figure}[t]
\includegraphics[width=.95\columnwidth]{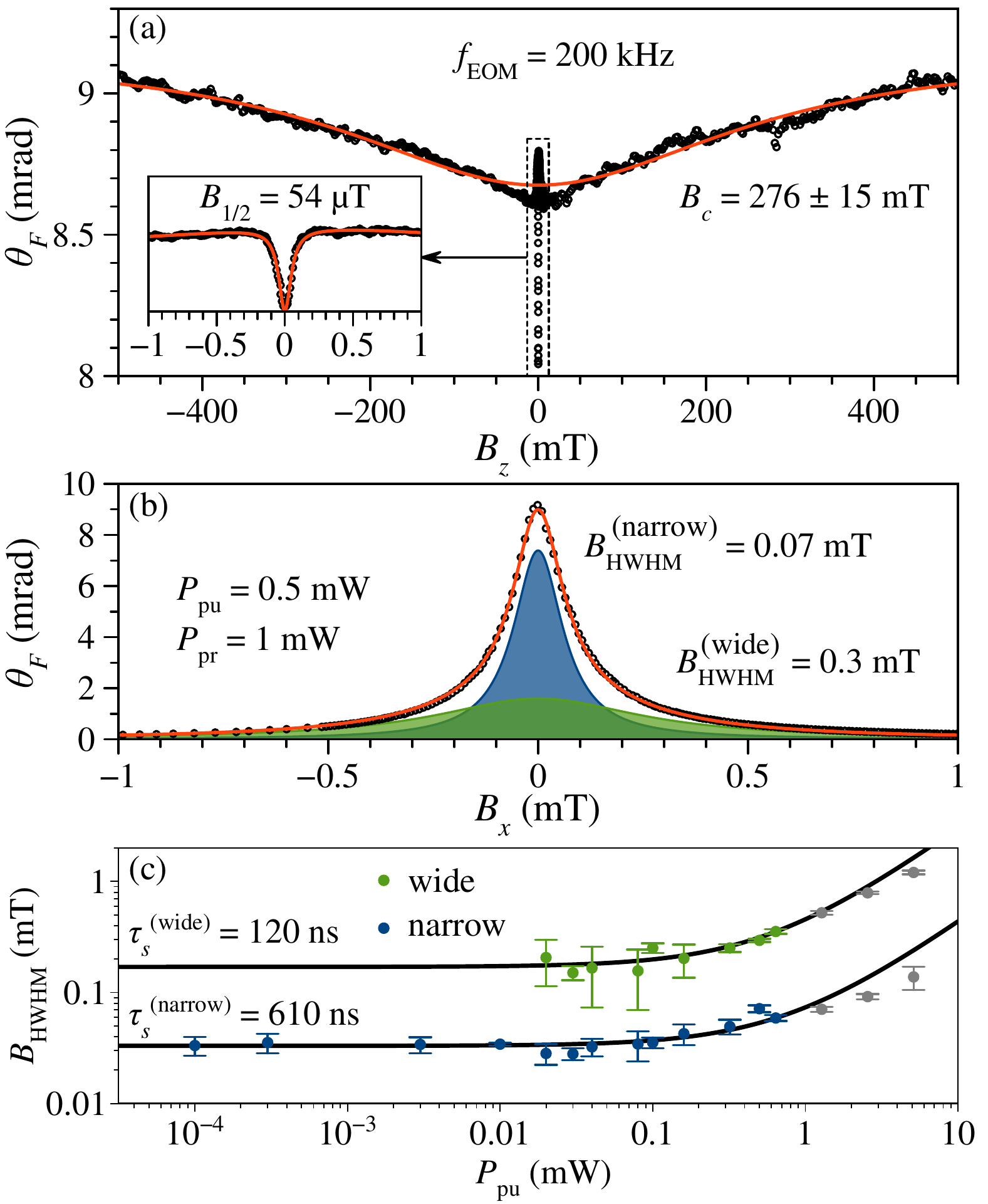}	
\caption{\label{fig:FR}(a) Spin polarization recovery in longitudinal magnetic field (points) and its fit by $\theta_F(B_z) = \theta_\infty/[1+(\tau/\tau_c)(1+(B_z/B_c)^2)^{-1}]$ (red line). Insert is a close-up for the structure around zero field with a half-width at half-minimum $B_{1/2}=54$\,$\mu$T. Red line is a Lorentzian fit. (b) The Hanle depolarization curve (circles) and its fit by a bi-Lorentzian function (red line). Shaded curves show the decomposition with two characteristic spin relaxation times. (c) Pump power dependence for both extracted components. Gray colored data are not following a linear dependence due to saturation effects and are not considered. Black lines are linear fits.
}
\end{figure}

To characterize the electron spin relaxation dynamics, we first perform FR measurements. To avoid effects of nuclear spin polarization, the measurements are done using polarization modulation of the pump light by the electro-optic modulator switched between $\sigma^+$ and $\sigma^-$ light polarization at frequency $f_\mathrm{EOM} = 200$\,kHz. The FR of the probe beam is detected by a lock-in technique at the frequency of modulation. Application of the longitudinal $B_z$ field recovers the electron spin polarization while the transverse field $B_x$ erases it, as shown in Figs.~\ref{fig:FR}(a) and ~\ref{fig:FR}(b). Note that the widths of the polarization recovery curve (PRC) [Fig.~\ref{fig:FR}(a)] and the Hanle curve [Fig.~\ref{fig:FR}(b)] differ by more than three orders of magnitude. Interestingly, such a behavior is common for semi-insulating samples of GaAs:Si~\cite{SokolovPRB17}. From the dataset, we extract the half-width at half-minimum of the PRC, $B_c = 276 \pm 15$\,mT, corresponding to a correlation time $\tau_c = 75 \pm 5$\,ps. We additionally point out, that the insert in Fig.~\ref{fig:FR}(a) demonstrates a narrow dip structure, which is a result of competition between nuclear spin cooling and nuclear spin warm-up in the oscillating Knight field of the electrons~\cite{SokolovPRB17}.

Furthermore, the Hanle curve is best fitted using two Lorentzians. Their half-width at half-maximum at corresponding laser powers is: $B_\mathrm{HWHM}^{(\text{narrow})} = 0.07$\,mT and $B_\mathrm{HWHM}^{(\text{wide})} = 0.3$\,mT, see Fig.~\ref{fig:FR}(b). The presence of two components indicates the presence of two sets of electrons, contributing to this relaxation with the lifetimes of $T_s^{(\text{narrow})} = 286$\,ns and $T_s^{(\text{wide})} = 67$\,ns. Here we used:
\begin{align} \label{eq:relaxation_time_Hanle}
T_s = \frac{\hbar}{g_e \mu_B B_{\text{HWHM}}},
\end{align}
with $|g_e|=0.568$ being the electron $g$ factor (its value is taken from the spin noise measurements in magnetic field at the same sample position, see below), $\hbar$ is the reduced Planck constant, and $\mu_B$ the Bohr magneton. In general, the spin lifetime $T_s$ can be defined as: $1/T_s = 1/\tau_s + G/n_0$~\cite{crooker2009spin,HeisterkampPRB15}. Here, $G$ is the generation rate of carriers that is dependent on the power of the optical excitation, $n_0$ is the carrier concentration, and $\tau_s$ the intrinsic longitudinal spin-relaxation time.
Therefore, by extrapolation to zero pump power, it is possible to extract the $\tau_s$. Figure~\ref{fig:FR}(c) demonstrates such an experiment with $\tau_s^{(\text{narrow})} = 610 \pm 10$\,ns for the narrow component. The broad component disappears below $P_\text{pu}=0.02$\,mW and extrapolates to the value of $\tau_s^{(\text{wide})}= 120\pm 10$\,ns.

\begin{figure}[t]
\includegraphics[width=.95\columnwidth]{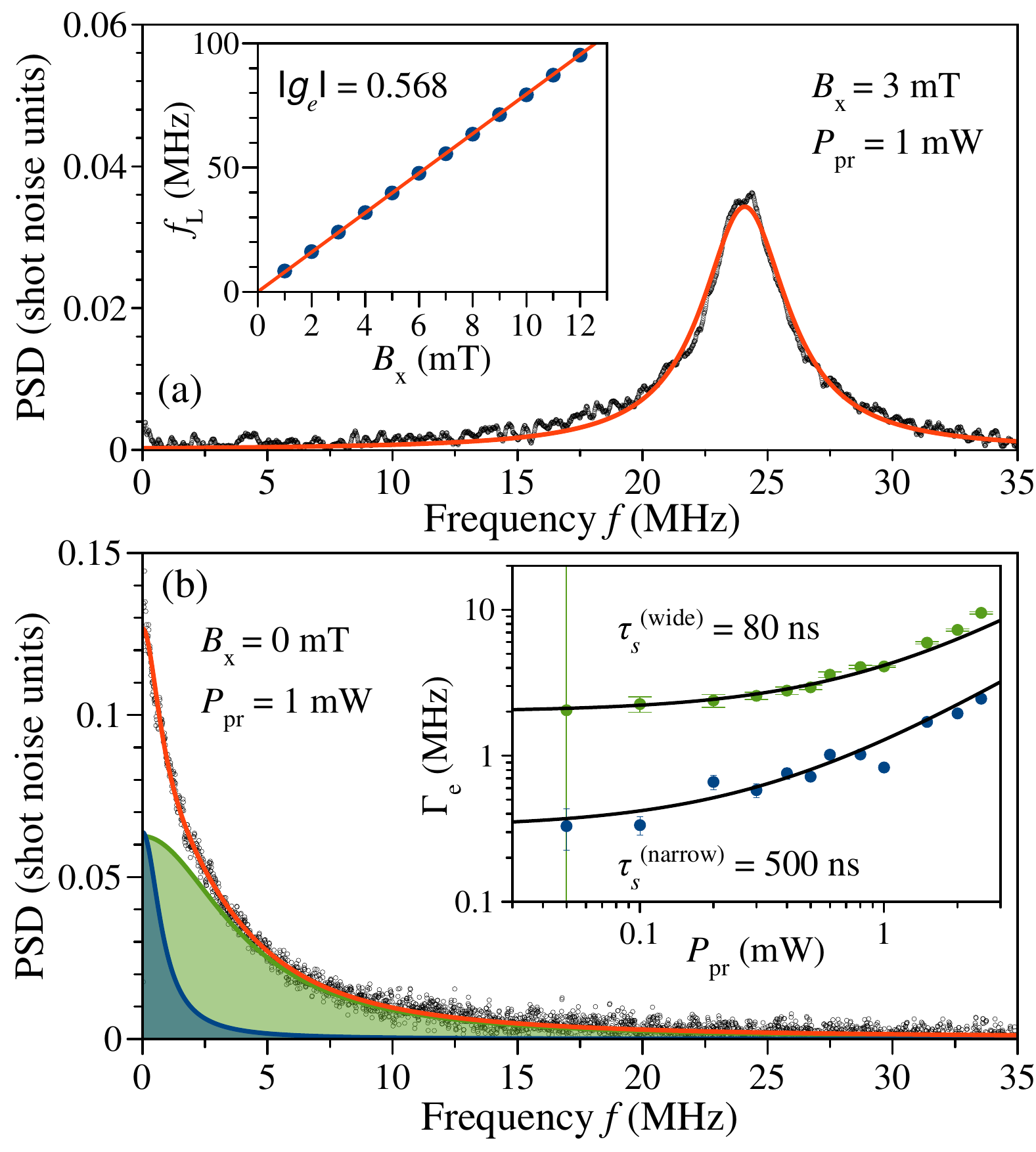}
\caption{\label{fig:SNS} (a) Example of a spin-noise spectrum measured for $B_\text{x} = 3$\,mT and $P_{\text{pr}}=1$\,mW. The power spectral density (PSD) is expressed in shot-noise units. Red solid line is a fit by a single Lorentzian function. Inset shows the dependence of the spin noise peak on $B_\text{x}$. Red line is a fit by Eq.~(\ref{eq:larmor_frequency}) with $|g_e|=0.568 \pm 0.001$. (b) Spin-noise spectra measured at $P_{\text{pr}}=1$\,mW and $B_\text{x}=0$\,mT. The data are fitted by a bi-Lorentzian function, red line. Shaded curves show the two fit components, narrow (blue) and wide (green), correspondingly. Inset shows the half width at half maximum ($\Gamma_e$) for each component versus probe power. Black curves show linear fits.}
\end{figure}

Next, to avoid unnecessary spin polarization of carriers by pumping, we use SNS. Here, the probe beam detects fluctuations of the electron-spin polarization in its ground state without optical pumping of the spin system~\cite{mullerSemiconductorSpinNoise2010, zapasskii2013spin}. In the usual case, the spontaneously appearing spin excitations decay exponentially in time and are, therefore, observed as a single Lorentzian peak in the spin-noise power-spectrum~\cite{sinitsyn2016theory, Smirnov2020_review}, see Fig.~\ref{fig:SNS}(a), measured at $B_\text{x} = 3$\,mT and $P_{\text{pr}}=1$\,mW. Following the dependence of the peak position ($f_\text{L}$) versus the external transverse magnetic field $B_\text{x}$, one can determine the Larmor $g$-factor $g_e$. The inset in Fig.~\ref{fig:SNS}(a) shows this dependence, characterized by the Larmor frequency $\omega_\text{L} = 2\pi f_\text{L}$. The red solid line is a fit by the equation:
\begin{align} \label{eq:larmor_frequency}
f_\text{L} = g_e \mu_{\text{B}} B_\text{x} / h,
\end{align}
with the Planck constant $h$.

The $B$-linear fit yields $|g_e| = 0.568 \pm 0.001$. The electron $g$~factor in InGaAs is expected to be slightly lower than in bulk GaAs with $g_{e,\,\text{GaAs}}\approx -0.44$~\cite{crooker2009spin}, as the additional indium content contributes with $g_{e,\,\text{InAs}} = -15$~\cite{oestreich1995temperature, sadofyev2002large} (one measures here the absolute value, but a negative sign is expected). Further investigations of the InGaAs sample indicate a spatial inhomogeneity of the electron $g$~factor across the sample, varying between $-0.53$ and $-0.6$, which can be related to a gradient of indium content in the epilayer.

To compare the SNS with the preceding Hanle measurements, we measure the probe power dependence of the SNS at $B_\text{x}=0$\,mT. Figure~\ref{fig:SNS}(b) demonstrates the observed peak centered at zero frequency which is best fitted with two Lorentzian, the sum of which is shown by the red curve. The HWHM of each peak, $\Gamma_e$, is inversely proportional to the electron-spin lifetime $T_s$ according to:
\begin{align} \label{eq:realxation_time}
T_s = \frac{1}{2 \pi \Gamma_e}.
\end{align}
In the optical SNS, the spin lifetime (or the peak width) is also affected by the non-zero probe power, in a similar way as in the Hanle effect. Experimentally, one can access the intrinsic spin-relaxation time by a power-dependent measurement with extrapolation to zero probe power, as presented in the inset of the Fig.~\ref{fig:SNS}(b). The intrinsic widths correspond to $\tau_s^{(\text{wide})} = 80 \pm 7$\,ns and $\tau_s^{(\text{narrow})} = 500 \pm 100$\,ns. These values compare well with the ones measured using the Hanle effect. The bigger error values are related to the limited sensitivity of the spin noise setup at lower probe power. The power range can be potentially extended to much smaller values using the homodyne detection demonstrated in Ref.~\cite{petrovIncreasedSensitivitySpin2018}.

The electron-spin lifetime of GaAs is well studied for various doping densities~\cite{DzhioevPRB02,BelykhPRB17,LonnemannPRB17}.
It depends strongly on the doping concentration and has a maximum right below the Mott-insulator transition (MIT) with $\tau_s \approx 800$\,ns for $n_D = 6.6 \times 10^{15}$\,cm$^{-3}$~\cite{LonnemannPRB17}. At doping densities around $n_D = 4.4 \times 10^{16}$\,cm$^{-3}$, close to the doping of the studied InGaAs epilayer, the relaxation time is found to be one order of magnitude shorter. The spin relaxation time in InGaAs could be further enhanced by optimizing the doping concentration.


In the absence of externally applied magnetic fields ($B_\text{x} = 0$\,mT), the spin-noise signal consists of a peak centered at frequency 0\,MHz, which describes the spin relaxation along the $z$-axis, see Fig.~\ref{fig:SNS}(b). This observation suggests that the spin noise is produced by electrons, which are either weakly affected by the surrounding nuclear spins due to motional narrowing, or the nuclear spin fluctuations cannot be considered as frozen and have a short nuclear spin correlation time ($\tau_c$)~\cite{MerkulovPRB02,glazov2012prb,glazovSpinNoiseLocalized2015,Smirnov2020_review}.
Such a nuclear spin dynamics could be driven by the Knight field of the electrons, or by the interaction of the nuclei quadrupole moments with strain and random electric fields in the structure~\cite{glazovElectronNuclearSpin2018}.

In the case of a strong coupling to the nuclei or of long $\tau_c$, one would expect to observe an additional peak at non-zero frequency, which would be related to the spin precession in the effective magnetic field produced by the random ‘‘frozen'' nuclear spin fluctuations with components orthogonal to the $z$-axis.

A corresponding observation was reported in Ref.~\cite{berskiInterplayElectronNuclear2015}, which discusses spin noise studies of a 10\,$\mu$m GaAs:Si epilayer with a low donor concentration of $n_D\approx 1\times 10^{14}$\,cm$^{-3}$. This represents a situation with a long correlation time, leading to a two-peak structure at zero external magnetic field.

On the opposite side, at higher donor concentrations ($n_D \approx  (1 - 7) \times 10^{16}$\,cm$^{-3}$), the spin noise is primarily produced by the free electrons fluctuating at the edge of the Fermi sea and no two-peak structure is observed at $B_\text{x} = 0$\,mT, see Ref.~\cite{crooker2009spin}. This is additionally supported by measurements of the temperature dependence demonstrating a linear increase of the integral spin-noise power with temperature.

\begin{figure}[t]
\includegraphics[width=.95\columnwidth]{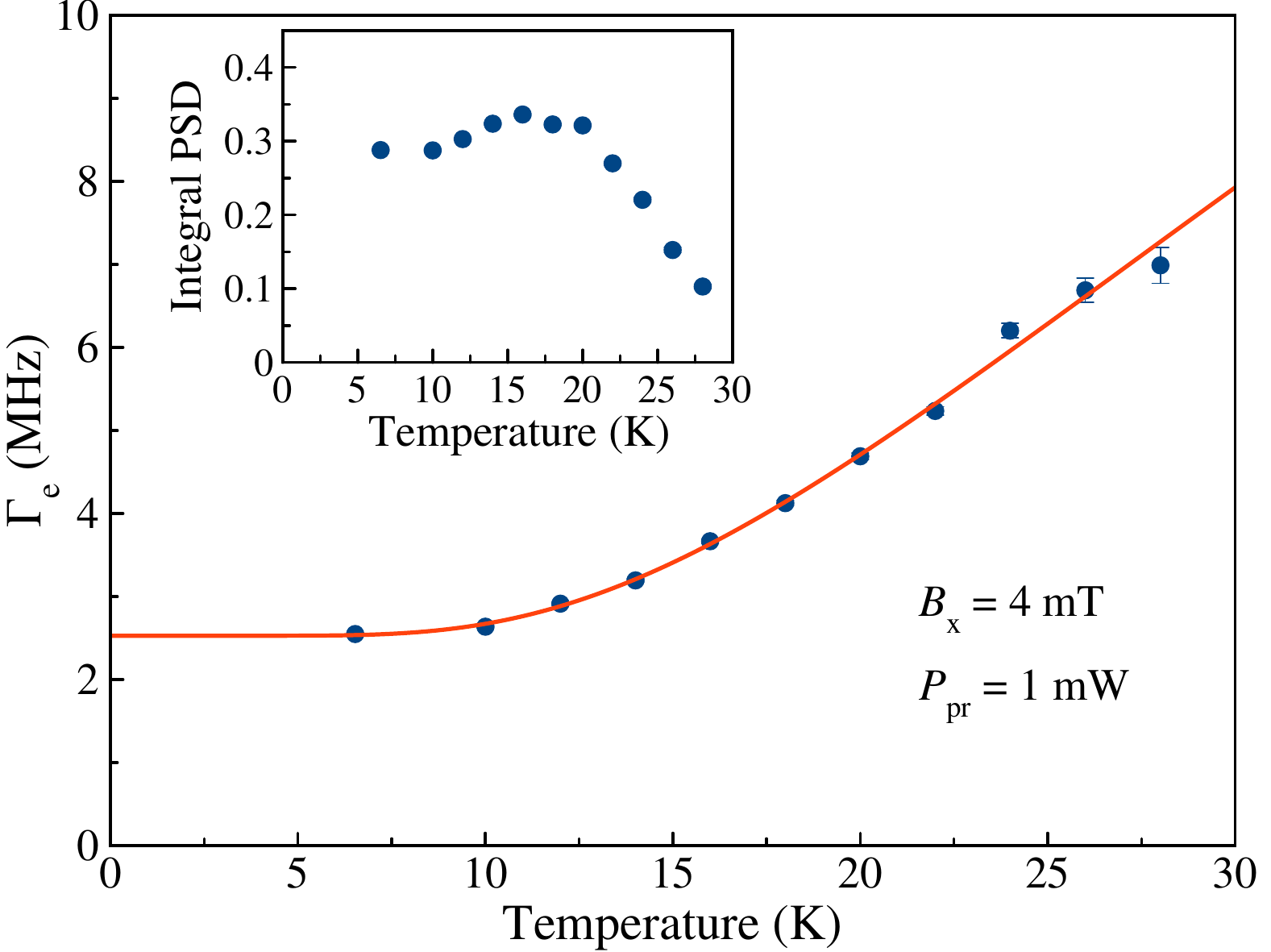}
\caption{\label{fig:arhenius} Temperature dependence of the spin-noise peak width $\Gamma_e$ measured at $B_\text{x} = 4$\,mT, using $P_\text{pr} = 1$\,mW. Red line is a fit by the Arrhenius equation, allowing us to determine the activation energy $E_\text{a}=4.7\pm 0.2$\,meV. Inset demonstrates the spectrally integrated noise power versus temperature.}
\end{figure}

Our observations demonstrate that the spin noise peak measured at magnetic fields above 1\,mT can be fitted well with a single Lorentzian, see Fig.~\ref{fig:SNS}(a). This observation still needs further investigations, but will be used here to simplify the measurement evaluation.
To gain more insight into the degree of localization of the electrons we conduct a temperature-dependent measurement of the spin-noise spectra for our sample. Figure~\ref{fig:arhenius} demonstrates the variation of the $\Gamma_e$ of the peak measured at $B_\text{x} = 4$\,mT.
The peak-width is constant up to the temperature of $\sim$10\,K and continuously increases above. The inset to the figure additionally demonstrates that the integral noise power remains about constant up to $T = 20$\,K and drops fast with further increase of the temperature. To describe it qualitatively, we use the Arrhenius equation: \looseness=-1
\begin{equation} \label{eq:arrhenius}
\Gamma(T) = \Gamma_g + \Gamma_{exc} \exp \left( -\frac{E_a}{k_\text{B}T} \right)
\end{equation}
with the activation energy, $E_a$, the relaxation rate $2\pi\Gamma_g = 15.9 \pm 0.1$\,$\mu$s$^{-1}$ of the ground state, and the relaxation rate $2\pi\Gamma_{exc} = 207 \pm 13$\,$\mu$s$^{-1}$ that characterizes the strength of carrier-phonon interaction~\cite{grundmann2010physics, greilich2012spin}. The measurement implies that the spin noise signal is produced by residual electrons that are localized at 6\,K, as the detected activation energy $E_a = 4.7 \pm 0.2$\,meV corresponds to a much higher temperature of about $E_a / k_\text{B} \approx 55$\,K. This activation energy agrees with typical values for GaAs:Si, where electrons are localized at donor centers~\cite{Bisotto_2005}.

We note here that the localized states, most probably, originate from pairs of closely situated donors screened by the degenerate electron gas~\cite{GiriPRL13}. The bound-carrier density is, at least, an order of magnitude smaller than $n_0$. An estimation yields~\cite{GiriPRL13} $N_b = (4/3)\pi\ell^3n_0^2\exp[-(4/3)\pi\ell^3n_0]$ where $\ell = (1/2)[\pi a_B^3/(3 n_0)]^{1/6}$ is the screening length and $a_B  =\hbar/\sqrt{2m_{\text{eff}} E_a} = 11$\,nm is the Bohr radius calculated using the effective mass $m_{\text{eff}} = 0.067 m_0$ ($m_0$ being the free electron mass in vacuum) and the activation energy $E_a = 4.7$\,meV, which is close to the standard Bohr radius for  GaAs:Si~\cite{OpticalOrientationBook}. This gives $N_b \simeq 4.2\times 10^{15}$\,cm$^{-3}$ for $n_0 = 3.9\times 10^{16}$\,cm$^{-3}$, measured at low $T$.

\section{\label{sec:depolarization}Nuclear-spin dynamics}

\subsection*{Decay of nuclear polarization}\label{subsec:decay}

In the previous section, we have discussed the effect of the unpolarized nuclear-spin bath on the localized electron spins. This can also be seen oppositely, the electron Larmor frequency can be used as a sensor for the effective magnetic fields in the localization area of the electron. Here, we use this sensor to study the relaxation dynamics of the polarized nuclear spins back to thermal equilibrium by the time-resolved version of SNS, as proposed by Ref.~\cite{smirnovPRB2015}. Again, the electron Larmor frequency is tested using SNS with a time-resolution of $\sim 1$\,s. This is chosen as a compromise between the required accumulation time for reliable peak detection and the shortest timescale of the observed nuclear-spin polarization decay.

We use the same technique as presented in Ref.~\cite{ryzhov2015measurements}. The measurement starts with dynamic nuclear-spin polarization by optically polarized electron spins, produced by the circularly polarized pump. Depending on the relative direction of the applied longitudinal magnetic field and the helicity of the circular polarization of the pump, one can determine the relative orientation of the nuclear spins and the electron spin polarization, or the nuclear spin temperature $\Theta_N$~\cite{OpticalOrientationBook}. In our experiments we used $B_\text{z} = 10$\,mT, $B_\text{x} = 0$\,mT. After the pumping period, the pump beam was blocked by a shutter, $B_\text{z}$ was set to zero, and $B_\text{x} = 4$\,mT was applied to detect the Larmor frequency with the probe laser. If the switching of the magnetic field happens adiabatically, the created nuclear polarization follows the direction of the external field~\cite{AbragamProctor}. At the same moment, the detection period was started.

The Larmor frequency determined from the peak position in the noise spectra is proportional not only to the transverse magnetic field, but also to the nuclear-spin polarization $p_N$ created by the optical pumping. It changes Eq.~(\ref{eq:larmor_frequency}) to:
\begin{align} \label{eq:frequency_overhauser_field}
f_\text{L}(t) = g_e \mu_{\text{B}} \bigl( B_\text{x} + b_{\text{max}}p_N(t) \bigr)/h.
\end{align}
The nuclear-spin polarization induces the Overhauser field $B_N(t) = b_{\text{max}} p_N(t)$ with a maximum value $b_{\text{max}}$. For InGaAs with 3\% of indium we estimate the maximal Overhauser field of $b_{\text{max}} = \sum_j I_j A_j \chi_j n_j/ g_e \mu_\text{B} = 4.125$\,T, with $I_j$ being the nuclear spin of the corresponding isotope, $A_j$ the hyperfine constant, $\chi_j$ its abundance, and $n_j$ the respective fraction of the nuclei in the material composition~\cite{MerkulovPRB02}. The electron $g$-factor is $|g_e|=0.6$, see below.

\begin{figure}[t]
    \includegraphics[width=.95\columnwidth]{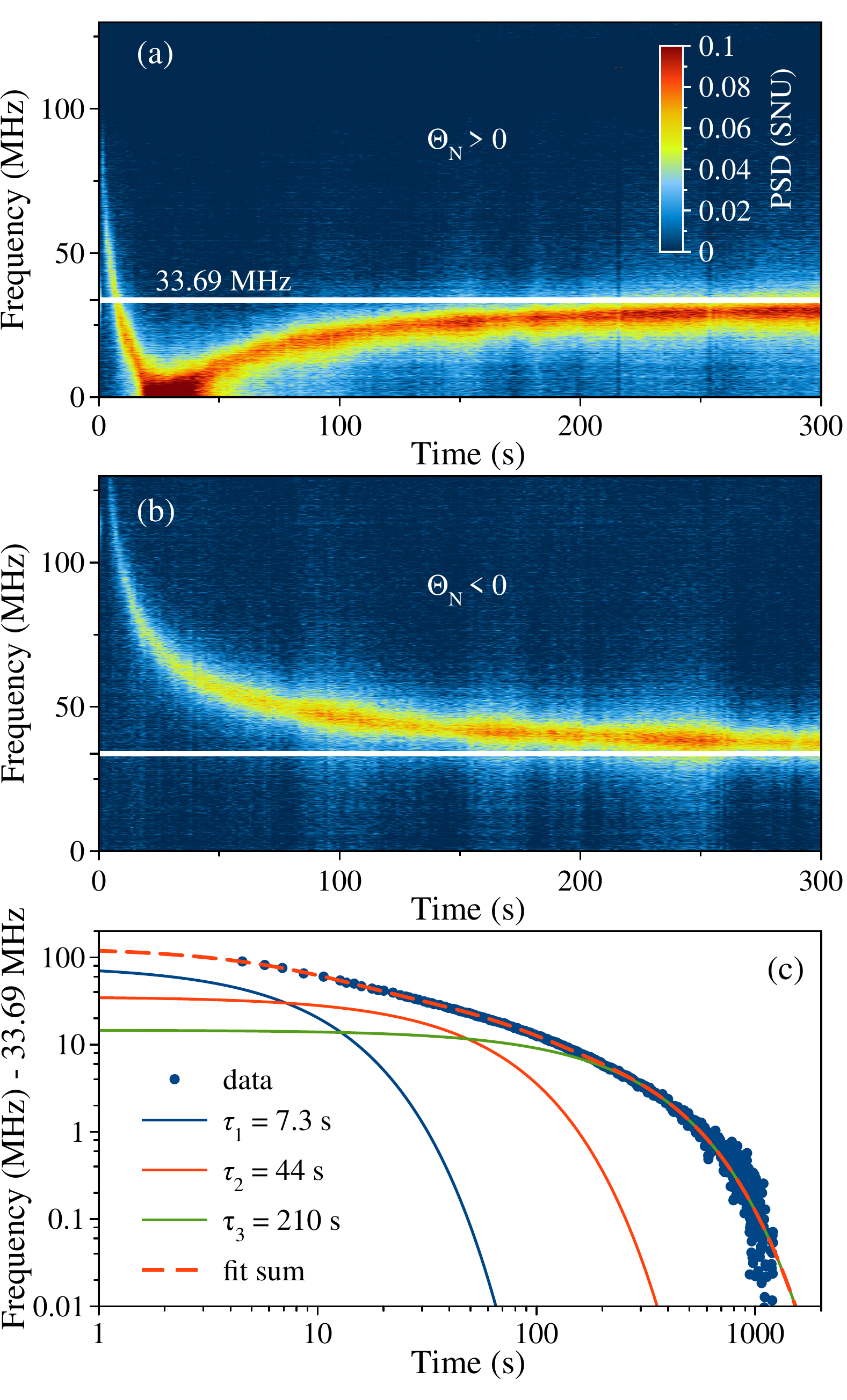}
\caption{\label{fig:nuclear_dynamics} Colormaps of time-resolved SNS after optical pumping for five minutes. They display the decay of the nuclear spin polarization (Overhauser field) for two different directions of the longitudinal field $B_\text{z}$ while pumping, leading either to a positive nuclear spin temperature $\Theta_N$ shown in (a) or to a negative $\Theta_N$ in (b). (c) The decay of the spin-noise peak position in (b) can be described by a multi-exponential function with three characteristic decay times $\tau_1, \tau_2$, and $\tau_3$. The frequency 33.69\,MHz shown by the white horizontal line, marks the peak position without nuclear polarization.}
\end{figure}

Figures~\ref{fig:nuclear_dynamics}(a) and \ref{fig:nuclear_dynamics}(b) show colormaps with the variation of the noise spectra versus observation time after pumping for five minutes with a pump power of $P_{\text{pu}} = 0.3$\,mW. In combination with a right circularly polarized pump, the induced Overhauser field points either in the direction opposite to the magnetic field $B_\text{x}$, see Fig.~\ref{fig:nuclear_dynamics}(a), or in the same direction, Fig.~\ref{fig:nuclear_dynamics}(b). The presented spin-noise spectra are taken every second with a probe power of $P_{\text{pr}} = 1$\,mW. The time dependence unveils the Overhauser field decay. When the decay is completed, the Larmor frequency remains constant at $f_\text{L} = 33.69$\,MHz corresponding to the externally applied field $B_\text{x} = 4$\,mT~\footnote{This frequency is determined by an additional measurement at $B_\text{x} = 4$\,mT without preceding nuclear polarization.}. This implies a $g$ factor $g_\text{e} = -0.6$ which indicates a different sample position compared to previous measurements but is well within the range of $g$ factors across the epilayer. Furthermore, the spectra at short times demonstrate a broadening of the noise peak due to the spatial inhomogeneity of the Overhauser field distribution. In the discussion below we concentrate on the case with negative nuclear spin temperature, see Fig.~\ref{fig:nuclear_dynamics}(b).

The decay of nuclear polarization is related to the interaction with the electron-spin system and to the dipole-dipole interaction between the nuclear spins~\cite{OpticalOrientationBook}. Additionally, increased quadrupolar effects in the studied sample are expected to influence the relaxation dynamics, see the next chapter~\cite{VladimirovaPRB18}. Depending on the dominating interactions, the decay of the Overhauser field can be described in a simplified way by using a sum of exponential functions with characteristic decay times $\tau_i$ and the corresponding amplitudes $a_i$. In the studied InGaAs epilayer, we explicitly detected three different relaxation times, referred to as $\tau_1$, $\tau_2$, and $\tau_3$, leading to:
\begin{equation}
p_N(t) = a_1 e^{-t/\tau_1} +  a_2 e^{-t/\tau_2} +  a_3 e^{-t/\tau_3}.
\end{equation}
The need for three decay times is illustrated in Fig.~\ref{fig:nuclear_dynamics}(c).
In this measurement, the shortest detected time is $\tau_1 = 7.3 \pm 0.1$\,s, the middle decay time is $\tau_2 = 44 \pm 1$\,s, and the longest one is $\tau_3 = 210 \pm 2$\,s.

\begin{figure}[t]
    \includegraphics[width=.95\columnwidth]{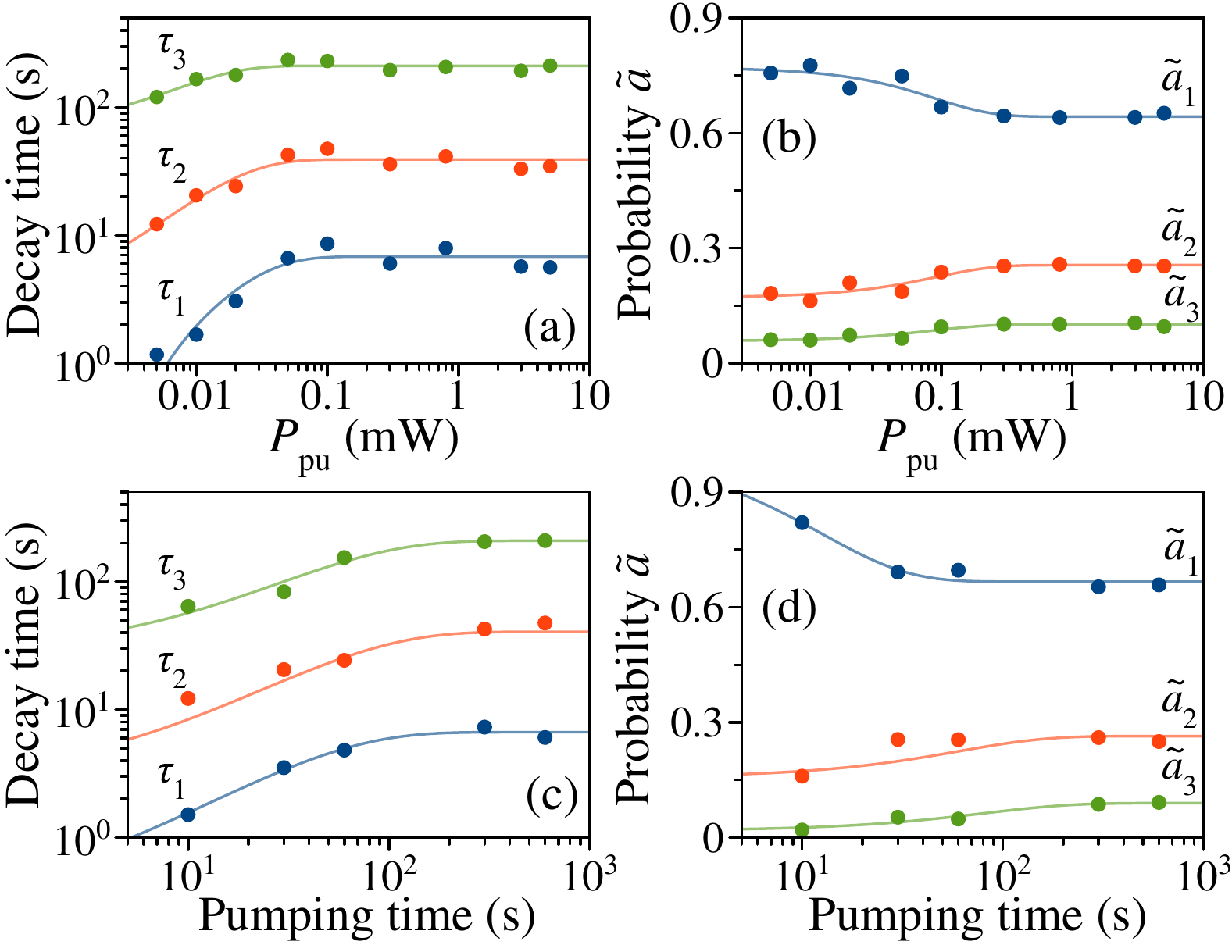}
\caption{\label{fig:power_time_dep} (a) Overhauser-field decay time versus pumping power. Varying pump powers between $P_\text{pu} = 5$\,$\mu$W and $P_\text{pu} = 5$\,mW lead to different decay time constants of the Overhauser field. Pumping time is fixed at 10\,min. (b) Variation of the decay probability components with pumping power. (c) Dependence of the decay times on the pumping time for a fixed pump power $P_\text{pu} = 0.3$\,mW. (d) Variation of the decay probabilities with pumping time. Solid lines are exponential fits to the data.}
\end{figure}

To get more insight into the relaxation dynamics we conducted a series of measurements for varying pump power and pump time. For the pump power dependence, $P_{\text{pu}}$ was varied from $5$\,$\mu$W to $5$\,mW. During 10\,min of pumping with $P_{\text{pu}}$, the longitudinal magnetic field $B_\text{z} = 10$\,mT is applied.
The subsequent time-resolved spin noise spectroscopy measures the polarization decay with the probe beam applied at the transverse magnetic field $B_\text{x} = 4$\,mT.
The development of all three decay times in Fig.~\ref{fig:power_time_dep}(a) indicates a rise of times with increasing pump beam power until it exceeds the power of 0.1\,mW.
Above this power, the nuclear-spin polarization time saturates. Below $P_\mathrm{pu} = 0.1$\,mW, the decay times decrease exponentially as shown by the solid line fits in Fig.~\ref{fig:power_time_dep}(a).
Above the critical pump power, between 0.1\,mW and 5\,mW, the decay times remain relatively stable with averaged values of
\begin{subequations}
\begin{align}
\tau_1 &=  7 \pm 1\,s, \\
\tau_2 &=  38 \pm 6\,s, \\
\tau_3 &=  210 \pm 10\,s.
\end{align}
\end{subequations}
Figure~\ref{fig:power_time_dep}(b) shows the same exponential behavior and the same critical pump power for the decay probabilities. The normalized amplitude $\tilde{a}_i=a_i/(a_1+a_2+a_3)$ refers to the probability that the $i$-th of the three decays occurs.

In principle, a reduction of the pumping time is expected to affect the nuclear-spin system in the same way as a decrease of the pump power.
To confirm this expectation, different pumping times (10\,s, 30\,s, 1\,min, 5\,min, and 10\,min) were used for a time-resolved spin-noise experiment with $P_{\text{pu}} = 0.3$\,mW at $B_\text{x} = 4$\,mT.
Figures~\ref{fig:power_time_dep}(c) and ~\ref{fig:power_time_dep}(d) demonstrate that the qualitative trends of the decay times $\tau_i$ and the probabilities $\tilde{a}_i$ resemble the results of the pump-power dependence.
The pump time-limit required for a saturated nuclear polarization is between 1\,min and 5\,min.
Above this pump time, at $P_\mathrm{pu} = 0.3$\,mW, the nuclear spin system is saturated.
Together with the pump power limit of $0.1$\,mW and optical pumping for 10\,min, the pump time limit is a useful result for further investigations using time-resolved SNS with optical pumping on the saturated nuclear-spin systems.

We relate our measurements to similar studies on GaAs:Si epilayers with comparable doping, for which only two exponents were observed~\cite{giriNondestructive2013,ryzhov2015measurements}. More specifically, the authors of Ref.~\cite{ryzhov2015measurements} found two decay times $\tau_1 = 30$\,s and $\tau_2 = 300$\,s for the dielectric phase of GaAs doping ($n_D = 2 \times 10^{15}$\,cm$^{-3}$) and only one exponent $\tau = 150$\,s for the metallic phase ($n_D = 4 \times 10^{16}$\,cm$^{-3}$).
The setup was similar to the one used by us, with pump times of 1\,min to 5\,min, at $B_\text{z} = 12$\,mT, and, most importantly, $B_\text{x} = 4$\,mT, as the values of the decay times can strongly depend on the chosen $B_\text{x}$~\cite{vladimirova2017nuclear}.
For the dielectric phase, the authors argued that the shorter time is associated with the electron-assisted spin depolarization of nuclei close to the donor centers and the longer time to spin diffusion governed by dipole-dipole interaction between nuclei far away from the donor centers, outside of the Bohr radius.
\looseness=0

Applying this time designation to our data, we can suggest that the $\tau_3$ component belongs to the nuclear spin diffusion outside the Bohr radius of the electron. It should take a rather long time and requires a rather strong pump power for it to be observed. Further, the presented pump-power and pump-time dependence of the Overhauser field's depolarization support the idea that the shortest time $\tau_1$ should arise from a different decay mechanism than the longer decay times. Therefore, we assign it to electron-assisted depolarization, requiring a shorter pump time and weaker pump power for it to be present, as the electron spin has a direct hyperfine coupling to the nuclear-spin bath. The additional nuclear-spin decay time ($\tau_2$) in the InGaAs epilayer has a similar behaviour as $\tau_3$ and might originate from depolarization through quadrupole effects that are strongly enhanced compared to GaAs due to the indium content. To provide an experimental proof for such an enhancement we determine the local field, induced by the fluctuating nuclear spins.

\subsection*{Measurement of the local field}
\begin{figure}[t]
    \includegraphics[width=.95\columnwidth]{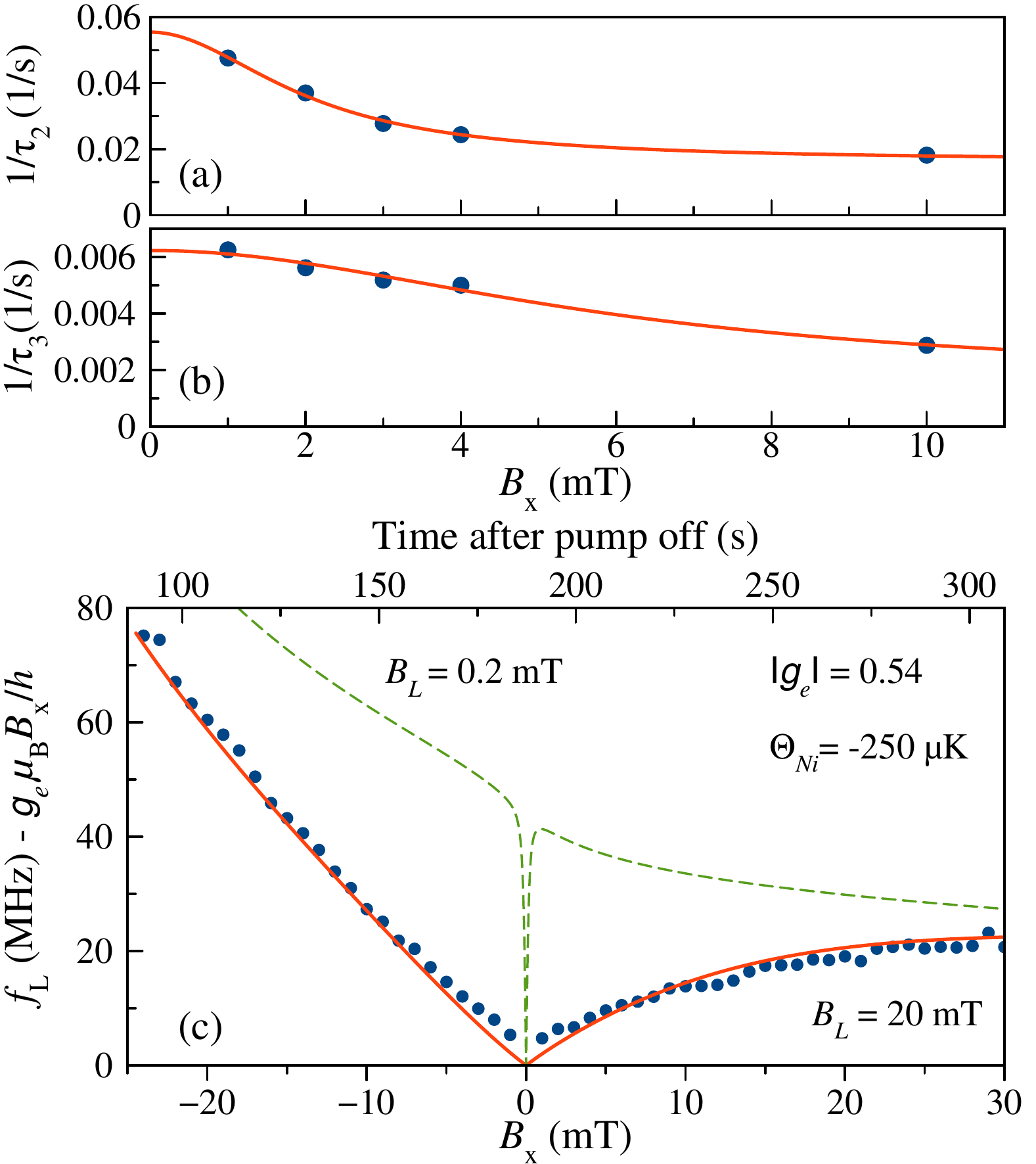}
\caption{\label{fig:BL} (a) and (b) Spin relaxation rates for the second and third time components measured as a function of the external magnetic field. Blue dots are data and red lines are fits by Lorentzian functions. (c) Overhauser field contribution to the frequency of the SN peak position for the adiabatic demagnetisation experiments. Blue dots are the experimentally determined frequencies after the subtraction of the electron Larmor frequency without nuclear contribution. Red line is the fit using the Eq.~\ref{eq:reMagnitization}, which gives the values of $B_{L}$ and $\Theta_{Ni}$. Green dashed line is an example of a fit with $B_{L}=0.2\,$mT. Top x-scale gives the time relative to the point of pump blocking.}
\end{figure}

One of the ways to describe the local field of interacting nuclear spins is to use the thermodynamics framework, particularly the concept of spin temperature and adiabatic demagnetization~\cite{AbragamProctor,goldman1970spin}. In that case, one introduces the spin temperature of the nuclear spin system $\Theta_N$. The nuclear spin polarization $p_N$ orients itself with external field $B$ and can be described by Curie's law $p_N = \gamma_N B C / \Theta_N$, with $\gamma_N$ being the nuclear gyromagnetic ratio and $C$ the Curie constant. An adiabatic decrease of $B$ from $B_i$ to $B_f$ would conserve the $p_N$, but would reduce the spin temperature by a factor of $B_f/B_i$. The local field sets the limit for that factor at low magnetic fields. The dipole-dipole interaction between nuclei determines the local field $B_L$, so $B_L\approx 0.2$\,mT for GaAs~\cite{paget1977low,dyakonov2008spin}. For nuclear isotopes having quadrupole moments, the local field can increase due to electrical fields or strain in the structure, which can be induced by lattice deformations~\cite{Litvyak21,Vladimirova2022PRB}. Once the external field $B \approx B_L$, the nuclear polarization is randomized, limiting the adiabatic spin cooling to $B_L/B_i$. However, once the external field is increased above $B_L$, the polarization recovers along the applied field direction. Therefore, to determine the $B_L$, it is required to measure the polarization $p_N$ of the nuclear spin system as a function of an external magnetic field $B_x$ by slowly ramping it through zero. For the nuclear spin polarization being optically prepared at a temperature $\Theta_{Ni}$ and $B_i > B_L$, we can describe the $p_N$ as~\cite{AbragamProctor,Kalevic2008}:
\begin{equation} \label{eq:spin_temp}
\begin{split}
p_N(t) = & \frac{B_x(t)}{3 k_B \Theta_N(t)} \hbar \langle\gamma_N(I+1)\rangle,\\
\frac{\Theta_N(t)}{\Theta_{Ni}} = & \frac{\sqrt{B_x^2(t)+B_L^2}}{\sqrt{B_i^2+B_L^2}},
\end{split}
\end{equation}
with $I$ being the nuclear spin and the angled brackets showing the averaging over the nuclear isotopes.

Further, we accommodate the experiments presented in Refs.~\cite{VladimirovaPRB18,Vladimirova2022PRB}. These references present a thorough description of a method to determine the $B_L$ applied to a $n$-doped GaAs epilayer, supported by the theoretical background. In our case, the sample was illuminated for 15 minutes with the pump beam of $P_\text{pu}=0.5\,$mW at $B_\text{z}=10$\,mT and $B_\text{x}=-30\,$mT. Then, the pump beam was blocked by a shutter, and the timer of the experiment was started. We waited for about one minute in darkness to allow for fast nuclear depolarization within the Bohr radius of electrons (the time scale of $\tau_1$, see Fig.~\ref{fig:nuclear_dynamics}(c)) and set the $B_\text{z}$ to zero. After that, the SN spectra are taken with one-second accumulation while stepping the $B_\text{x}$ from $-30$ to $30$\,mT with 1\,mT steps. To extract the evolution of the nuclear polarization $p_N$, we determined the peak position of the SN spectra at each magnetic field and subtracted the electron Larmor frequency ($g_e \mu_\text{B} B_x/h$) without nuclear polarization at the same field, compare with Eq.~\ref{eq:frequency_overhauser_field}. The value of the $g$-factor $|g_e|=0.54$ was determined independently at the same sample position without preceding nuclear polarization. The blue points in Fig.~\ref{fig:BL}(c) represent the extracted data. As discussed above, the nuclear polarization should recover once the magnetic field increases above zero. The observed asymmetry is related to the long accumulation times so that the spin-lattice relaxation reduces the signal, see the top x-scale in Fig.~\ref{fig:BL}(c) for the times after the pump blocking.
In Appendix~\ref{app:remag}, we consider the situation when remagnetization takes place together with spin-lattice relaxation. On the experimental side, it requires the knowledge of the magnetic field dependence of the spin relaxation. To establish that, we have done measurements similar to the one presented in Fig.~\ref{fig:nuclear_dynamics} for different fixed magnetic fields $B_x$. Figures~\ref{fig:BL}(a) and \ref{fig:BL}(b) represent the extracted values for the rates of the second and third fitting components $\tau_2$ and $\tau_3$, respectively. A Lorentzian function gives the best fit for these rate dependencies on the external magnetic field. It allows us to obtain the functional dependence of the relaxation rates on $B_x$~\cite{vladimirova2017nuclear}.

Finally, the red line in Fig.~\ref{fig:BL}(c) represents a fit by Eq.~\eqref{eq:reMagnitization} with two free fitting parameters: $B_L$ and $\Theta_{Ni}$. Best fit gives the $B_L=20\pm1\,$mT and $|\Theta_{Ni}|=250\pm10\,\mu$K. For comparison, we show the curve for $B_L=0.2\,$mT by the dashed green line~\footnote{We have additionally done a fine step (step size of 0.02\,mT) magnetic field measurements close to zero field to test for the presence of the narrow dip structure with $B_L\approx 0.2\,$mT. No appearance was observed.}. As one can see, the value of $B_L$ is two orders of magnitude larger than that measured for $n$-doped GaAs epilayers~\cite{VladimirovaPRB18}. In Refs.~\cite{VladimirovaPRB18,Vladimirova2022PRB}, it is also demonstrated that the quadrupole effects are responsible for an increase of $B_L$, leading to values of $B_L\approx 2$\,mT.

Such significant values of $B_L$ in our sample indicate a strong quadrupolar contribution to the local fields but also raise a concern about the validity of the spin temperature approach, compare with Ref.~\cite{Maletinsky2009}. To test it, we have done two additional experiments. At the optical pumping stage, $B_x$ was fixed now at $+30\,$mT in addition to $B_z=10\,$mT. Then, after the pump blocking and waiting for one minute, the field is swept to $B_x=-30\,$mT at a rate of (i) 320\,mT/s or (ii) 6\,mT/s. Once $B_x=-30$\,mT is reached, the experiments and data processing are done similarly. This resulted in values of $B_L$ and $\Theta_{Ni}$ that are identical within the error bounds to the case without a sweep, Fig.~\ref{fig:BL}(c). It indicates that the spin temperature approach is still valid in our case.

\subsection*{Spin diffusion model}

To advance our understanding of the nuclear-spin relaxation, we additionally analyze the decay of the nuclear polarization using the diffusion model, presented in Ref.~\cite{paget1982optical}. To do that, we calculate the diffusion equation:
\begin{equation}
\begin{split}
    \frac{dp_N(t,r)}{dt} = D \Delta p_N(t,r) - p_N(t,r)\left(T^{-1}_{1e}(r)+T^{-1}_{1,K}\right) +\\ G(t,r).
    \label{eq_diffusion}
\end{split}
\end{equation}
Here, $p_N(t,r)$ is the nuclear-spin polarization at distance $r$ from the center measured at time $t$, $T_{1e}(r) = T_{1e}(0)\, \exp(4r/a_\mathrm{B})$ is the position-dependent nuclear-spin relaxation time due to interaction with bounded electrons, $T_{1,K}$ is the time characterizing the nuclear-spin relaxation due to interaction with the Fermi-edge electrons (the Korringa mechanism~\cite{vladimirova2017nuclear}), $D$ is the nuclear-spin diffusion constant, and $G(t,r)$ is the pumping rate. The nuclear-spin relaxation rate at the donor origin is defined by
\begin{equation}
    \frac{1}{T_{1e}(0)} = \Gamma_t \Omega^2 \frac{2\tau_c}{1+\omega^2\tau_c^2}.
    \label{eq_T1e}
\end{equation}
Here, $\omega = g_e \mu_\mathrm{B} B_\text{x}/\hslash$ is the magnetic field given in frequency units, $\Gamma_t$ is the probability of occupation of the donor (for simplicity, we take $\Gamma_t = 1$), $\tau_c$ is the correlation time measuring the residing time of the electron at the donor interacting with nuclear spins with the magnitude given by:
\begin{equation}
    \Omega = \frac{A_\mathrm{hf}}{2\hslash}\frac{v_0}{\pi a_\mathrm{B}^3}
    \label{eq_Omega}
\end{equation}
where $A_\mathrm{hf} = 46$\,$\mu$eV is the electron-nuclear hyperfine constant averaged over the atom species in a unit cell, $v_0 = (0.283)^3$\,nm$^3$ is the two-atom unit-cell volume.

\begin{figure}[t]
    \includegraphics[width=.95\columnwidth]{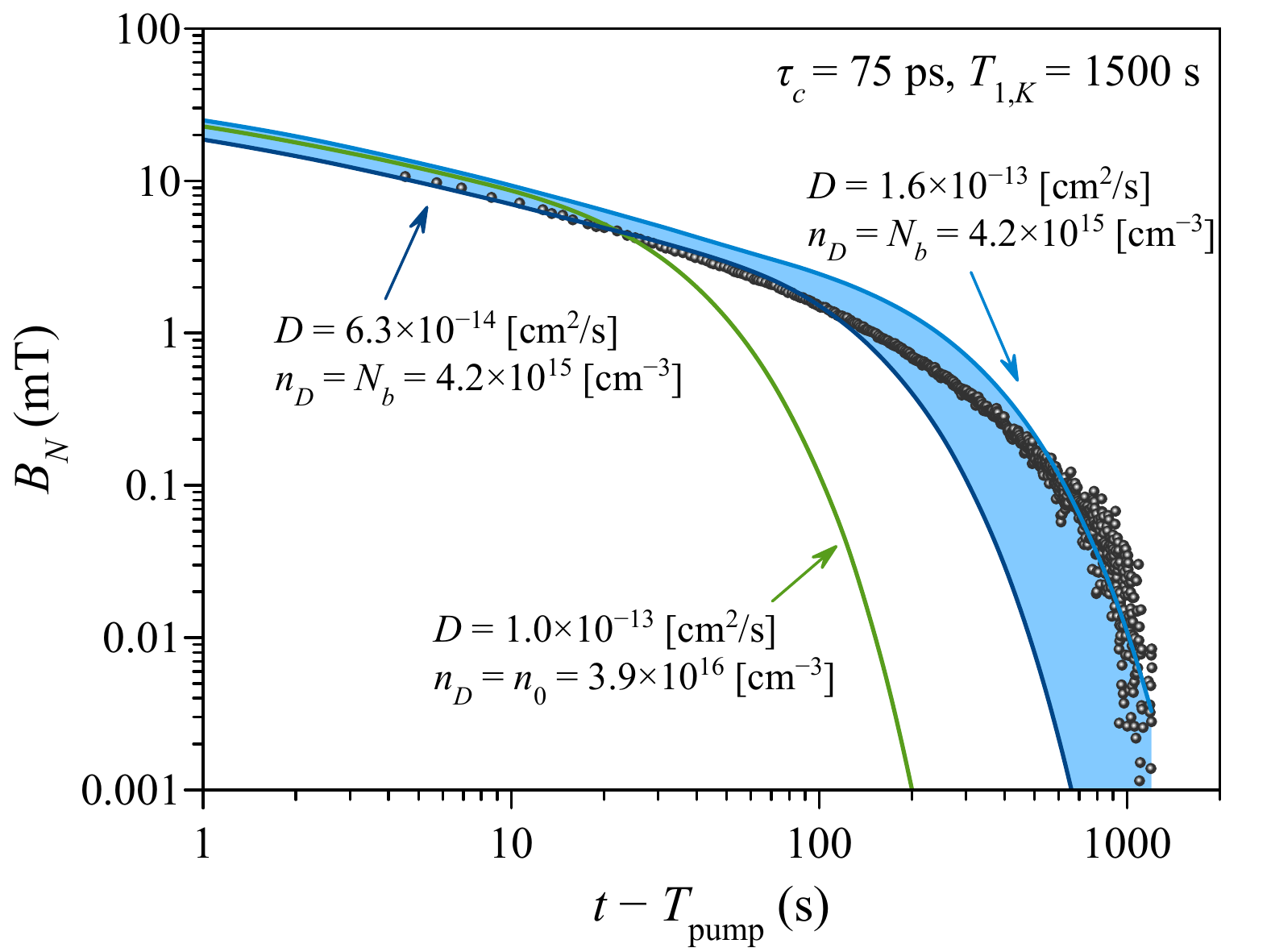}
\caption{\label{fig:nuclear_diffusiondecay} Decay dynamics calculated using Eqs.~\eqref{eq_diffusion} and \eqref{eq_BN} (solid lines) for various $n_D$ and $D$, $\tau_c = 75$\,ps estimated from the polarization recovery curve, and $T_{1,K} = 1500$\,s estimated for the doping density $n_0$. Circles show the experimental data retracted from Fig.~\ref{fig:nuclear_dynamics}(c).}
\end{figure}

Equation~\eqref{eq_diffusion} has no simple analytical solution, therefore we treat it numerically by substituting $p_N(t,r) = (1/r)\mathcal{P}_N(t,r)$ for numerical stability. The pumping rate is given by:
\begin{equation}
    G(t,r) = P_e\frac{I+1}{S+1}\frac{1}{T_{1e}(r)}\left[\Theta(t)-\Theta(t-T_\mathrm{pump})\right],
    \label{eq_G}
\end{equation}
where $S = 1/2$ and $I = 3/2$ are the electron and nuclear spins, $P_e = \langle S_z\rangle/S$ is the electron-spin polarization when pumping, and the term in brackets represents the switch-on and switch-off pumping at the moments $t = 0$ and $t=T_\mathrm{pump}$, given by theta-functions.
The calculation is performed in a region $R = n_D^{-1/3}$ for a given donor concentration with first and second-type boundary conditions at $r = 0$ and $r = R$, respectively.
To perform a direct comparison of the calculations with the experiment, the time evolution of the Overhauser field is calculated:
\begin{equation}
    B_N (t) = b_n\int_0^R \mathcal{P}_N(t,r)\,\exp\bigl(-2r/a_\mathrm{B}\bigr)\, r dr
    \label{eq_BN}
\end{equation}
where $b_n$ is a scaling factor. The results of the calculations for the spin relaxation at times $t-T_\mathrm{pump}$ are shown in Fig.~\ref{fig:nuclear_diffusiondecay}. As one can see from the figure, the experimental data retracted from Fig.~\ref{fig:nuclear_dynamics}(c) are reasonably well reproduced by the modeling, especially in the limit of short and long times.

However, this becomes only possible if the concentration of donors is reduced by an order of magnitude, down to $n_D = N_b = 4.2 \times 10^{15}$\,cm$^{-3}$. A relatively small variation of the diffusion constant $D$ allows one to fit better the experimental dependence at long times. This provides an estimation of $D$ when $n_D$ is known.
On the contrary, the calculations done for a nominal concentration $n_D = n_0 = 3.9\times10^{16}$\,cm$^{-3}$ provide a much faster Overhauser-field relaxation dynamics than the one observed experimentally (see green curve in Fig.~\ref{fig:nuclear_diffusiondecay}).
Note that, in principle, the quadrupole interaction could retard the spin diffusion, which would result in a reduced $D$ for fitting the data.
However, we find that a slightly higher value of $D$ is required to properly fit the experimental data at longer times than the one at the initial relaxation stage (see blue lines in Fig.~\ref{fig:nuclear_dynamics}).
We also find a small relative sensitivity of the model to a variation of $\tau_c$ at small times $t$. Furthermore, in our modeling, $\omega\tau_c \ll 1$ because the external field $B$ and the observed $B_N$ are small. In other words, the transition from the short to the long correlation-time regime, where $T_{1e}$ is affected most by $\omega\tau_c$~\cite{MerkulovPRB10}, is not achieved.

\section{\label{sec:concluction}Conclusion}

To conclude, we have investigated the influence of the indium contribution in the GaAs matrix on the donor-bounded carrier- and nuclear-spin relaxation dynamics. Besides the red shift of the band gap and donor emission energies we observe an enhancement of the carrier localization for comparable doping concentrations of the GaAs:Si structures. The dynamics of nuclear-spin relaxation reveals a complex three-exponential decay of the Overhauser field, which we interpret as result of the enhanced quadrupole effects induced by the indium. We provide experimental evidence for this enhancement by measuring the local field, which exceeds the quadrupole-free local field given by the dipole-dipole interaction by two orders of magnitude. This value of the local field puts the studied structure in the range between the low-stressed GaAs epilayers with $B_L\approx 2\,$mT, where the spin temperature approach is valid, and highly-stressed self-assembled InGaAs QD structures with $B_L = 300$\,mT~\cite{Maletinsky2009}, where the spin temperature approach breaks down. The modeling of the nuclear-spin polarization relaxation by the diffusion model suggests that the donor concentration should be reduced by an order of magnitude in order to match the experimental results. The origin of this discrepancy is not completely clear, but could be related to donor depletion due to surface charges, weak carrier localization, and the need to include the quadrupole interaction into the diffusion model. Further studies including the optimization of the silicon doping as well as an extension of the diffusion model are planned. Additionally, application of external stress to control the quadrupole interaction could be an option to influence the spin relaxation dynamics and, therefore, provide silicon-doped InGaAs epilayers as an advantageous alternative to GaAs:Si for low-temperature nuclear-spin ordering and to InGaAs QDs for reduced quadrupole effects.

\begin{acknowledgments}
We thank D.~S.~Smirnov and V.~V.~Belykh for fruitful discussions.
The sample was grown by using the facilities of the SPbU Resource Center ``Nanophotonics''. The transport measurements were performed at the P.\,N.\,Lebedev Physical Institute shared facility center.
The optical investigations presented in this work were done at TU Dortmund.
The data analysis and representation were performed by using the MagicPlot software. We acknowledge the financial support by the Deutsche Forschungsgemeinschaft in the frame of the International Collaborative Research Center TRR 160 (Projects A5 and A6) and the Russian Foundation for Basic Research (Grant No.\,19-52-12054 and 19-52-12043).
MYP and KVK acknowledge Saint Petersburg State University for the research grant 91182694.
\end{acknowledgments}

\appendix
\section{Remagnetization with spin-lattice relaxation}\label{app:remag}

Equation~\eqref{eq:spin_temp} is usually obtained from thermodynamic considerations, using entropy balance at equilibrium. That approach does not allow one to take into account relaxation processes. In order to generalize Eq.~\eqref{eq:spin_temp} for the case of slow (as compared to spin-spin processes) spin-lattice relaxation, we re-derive it from the condition of energy balance during a change of the magnetic field.
The rate of the energy change of the spin system under the action of a changing external magnetic field is:
\begin{equation} \label{eq:d_energy}
\frac{dE}{dt}=\frac{\partial\langle H \rangle}{\partial t}=-\langle \vec M \rangle \frac{\partial \vec B}{\partial t} = -\text{Tr}(\hat{M}_B^2)\beta B \frac{\partial B}{\partial t},
\end{equation}
with $\beta=(k_B \Theta_N)^{-1}$, $\hat{M}_B$ being the projection operator of the total magnetic moment of the nuclei on the external field $B$.
This equality is true, since the external magnetic field is the only parameter of the Hamiltonian of the spin system that depends on time directly~\cite{LandauLifshitz80}. On the other hand, the energy of the spin system is:
\begin{equation} \label{eq:energy}
\begin{split}
E & =\langle \hat{H}\rangle=-\beta \text{Tr}(\hat{H}^2)=-\beta(\text{Tr} \hat{H}_Z^2 + \text{Tr} \hat{H}_{SS}^2)\\
& =-\beta \text{Tr}M_B^2(B^2+B_L^2),
\end{split}
\end{equation}
with $\hat{H}_Z=-M_B B$ being the Hamiltonian of Zeeman interaction, $\hat{H}_{SS}$ the Hamiltonian of spin-spin interactions, and $B_L\equiv \sqrt{(\text{Tr}M_B^2)^{-1}\text{Tr}(\hat{H}_{SS})}$ is, by definition, the local field~\cite{goldman1970spin}, characterizing the strength of spin-spin interactions.
Accordingly, the total time derivative of the energy, considered as a function of the inverse spin temperature $\beta$ and the external magnetic field $B$, is:
\begin{equation} \label{eq:d_energy_full}
\begin{split}
\frac{dE}{dt} & =\frac{d}{dt}\left[-\beta(t)\text{Tr}M_B^2\left(B_L^2+B^2(t)\right)\right]=\\
& =-\text{Tr}M_B^2\left[\left(B_L^2+B^2(t)\right)\frac{d\beta(t)}{dt}+2\beta(t)B(t)\frac{dB(t)}{dt}\right].
\end{split}
\end{equation} 		
Setting Eqs.~\eqref{eq:d_energy} and~\eqref{eq:d_energy_full} to be equal, we obtain:
\begin{equation} \label{eq:equality}
\left(B_L^2+B^2(t)\right)\frac{d\beta(t)}{dt}+\beta(t)B(t)\frac{dB(t)}{dt}=0,
\end{equation}
that gives a differential equation for $\beta$:
\begin{equation} \label{eq:d_beta}
\frac{d\beta(t)}{dt}=-\beta(t)\frac{B(t)}{B_L^2+B^2(t)}\frac{dB(t)}{dt},
\end{equation}
which, in order to account for spin-lattice relaxation, should be complemented by a relaxation term of the form $-(\beta-\beta_L)/T_1$, where $\beta_L=(k_BT)^{-1}$ is the inverse temperature of the lattice, and the spin-lattice relaxation time $T_1$ depends, in general, on the magnetic field~\cite{vladimirova2017nuclear}. In experiments with optical cooling of nuclei in weak magnetic fields, the nuclear temperatures usually do not exceed a few millikelvin (otherwise there is no noticeable nuclear magnetization), and the lattice temperature is several kelvins. Therefore, we can set $\beta_L$ equal to zero, and the equation for $\beta$ takes a simple form:
\begin{equation} \label{eq:d_beta_simple}
\frac{d\beta(t)}{dt}=-\beta(t)\frac{B(t)}{B_L^2+B^2(t)}\frac{dB(t)}{dt}-\frac{\beta(t)}{T_1(B)},
\end{equation}
permitting an analytic solution. By dividing both sides of Eq.~\eqref{eq:d_beta_simple} by $\beta$ it is brought to the form:
\begin{equation} \label{eq:d_log_beta}
\frac{d\text{ln}\beta(t)}{dt}=-\frac{d}{dt}\text{ln}\sqrt{B_L^2+B^2(t)}-\frac{1}{T_1(B)}.
\end{equation}
If at the time moment $t=0$ the magnetic field was equal to $B_i$, and the inverse spin temperature equal to $\beta_i$, solution of Eq.~\eqref{eq:d_log_beta} yields the following time dependence of $\beta$:
\begin{equation} \label{eq:d_beta_final}
\frac{\beta(t)}{\beta_i}=\sqrt{\frac{B_L^2+B_i^2}{B_L^2+B^2(t)}}\exp\left(-\int_0^tT_1^{-1}(B(t')dt')\right).
\end{equation}

In the absence of spin-lattice relaxation, Eq.~\eqref{eq:d_beta_final} gives an expression for the inverse spin temperature under an adiabatic change in the magnetic field, usually obtained from the condition of entropy being constant in the adiabatic process~\cite{goldman1970spin}. In view of:
\begin{equation} \label{eq:MB}
\langle M_B \rangle =\text{Tr}\left(\rho_N\hat{M}_B\right)\approx\beta B \text{Tr}(\hat{M}_B^2),
\end{equation}
with $\rho_N$ being the density matrix for an ensemble of spins in high-temperature approximation:
\begin{equation} \label{eq:rho}
\begin{split}
\rho_N = & \frac{\exp\left[-\beta\left(\hat{H}_{SS}+\hat{H}_Z\right)\right]}{\text{Tr}\left(\exp\left[-\beta\left(\hat{H}_{SS}+\hat{H}_Z\right)\right]\right)}\approx\\
\approx & 1-\beta\left(\hat{H}_{SS}+\hat{H}_Z\right),
\end{split}
\end{equation}
we obtain the expression for the magnetization:
\begin{equation} \label{eq:Magnitization}
\begin{split}
\frac{\langle M_B(t)\rangle}{\langle M_B(0)\rangle}= & \frac{B(t)}{B_i}\sqrt{\frac{B_L^2+B_i^2}{B_L^2+B^2(t)}}\times\\
& \times \exp\left(-\int_0^tT_1^{-1}(B(t')dt')\right).
\end{split}
\end{equation}

In our experiments, a multi-exponential relaxation is observed, which is presumably related to spin diffusion in presence of spatially inhomogeneous spin-lattice relaxation due to hyperfine and quadrupole interactions~\cite{vladimirova2017nuclear}. The rigorous way to take these processes into account would be complementing Eq.~\eqref{eq:d_beta_simple} with a diffusion term and considering the spin temperature and $T_1$ as functions of coordinates. Here we used a simplified approach, assuming, that the contributions to the observed nuclear spin polarization, characterized by the spin-lattice times $\tau_i$ and amplitudes $a_i$, arise from different spin populations. This way, we obtain the following expression for the nuclear spin polarization under adiabatic remagnetization:
\begin{equation} \label{eq:reMagnitization}
\begin{split}
\frac{p_N(t)}{p_N(0)}= & \frac{B(t)}{B_i}\sqrt{\frac{B_L^2+B_i^2}{B_L^2+B^2(t)}}\times\\
& \times \sum_i a_i \exp\left(-\int_0^t \tau_i^{-1}(B(t')dt')\right).
\end{split}
\end{equation}
\vspace{0.5ex}


%

\end{document}